\documentclass[aps,prc,twocolumn,amsfonts,showpacs,floats,superscriptaddress,preprintnumbers,byrevtex]{revtex4}
\usepackage{epsfig}
\usepackage{amssymb} 
\newcommand{\nn}{\nonumber}
\newcommand{\nuc}[2]{$^{#1}${#2}}
\newcommand{\etal}{\emph{et al.}}
\begin{document}
\title{Configuration Mixing of\\ 
       Angular Momentum Projected Self-Consistent Mean-Field States\\ 
       for Neutron-Deficient Pb Isotopes}
\author{M. Bender}
\affiliation{Service de Physique Nucl\'eaire Th\'eorique,
             Universit\'e Libre de Bruxelles, C.P. 229, B-1050 Bruxelles,
             Belgium}
\author{P. Bonche}
\affiliation{Service de Physique Th\'eorique,
             CEA Saclay, 91191 Gif sur Yvette Cedex,
             France}
\author{T. Duguet}
\affiliation{Physics Division, Argonne National Laboratory, 
             Argonne, IL 60439}
\author{P.-H. Heenen}
\affiliation{Service de Physique Nucl\'eaire Th\'eorique,
             Universit\'e Libre de Bruxelles, C.P. 229, B-1050 Bruxelles,
             Belgium}
\date{November 23 2003}
%
%
\begin{abstract}
We study the low-lying collective excitation spectra of the 
neutron-deficient lead isotopes \nuc{182-194}{Pb} by performing a
configuration mixing of angular-momentum and particle-number 
projected self-consistent mean-field states. The same Skyrme 
interaction SLy6 is used supplemented by a 
density-dependent zero-range pairing force. This study supports 
the interpretation
of spectra made on the grounds of more schematic models in terms
of coexisting spherical, oblate, prolate and superdeformed 
prolate structures. The model qualitatively reproduces the
variation of the spectra with neutron number. Our results for $E0$ and $E2$
transition probabilities are compared with the few existing experimental 
data. Finally, we predict the presence of superdeformed bands at
low excitation energy in the most neutron-deficient isotopes. 
\end{abstract}
\pacs{21.60.Jz, 
      21.10.Pc, 
      27.70.+q, 
      27.80.+w  
}
\maketitle
%
%
\section{Introduction}
The neutron-deficient lead isotopes have been the subject of intense
experimental studies for nearly 20 years \cite{Jul01}. This continuous 
interest is mainly motivated by their rich excitation spectra
which strongly depend on the neutron number. 
First evidence for low-lying $0^+$ states in \nuc{192-198}{Pb} was
obtained from $\beta^+$-decay and electron capture of adjacent Bi 
isotopes~\cite{Dup84,Dup87}. At least one low-lying
excited $0^+$ state has now been observed in all even-even Pb isotopes
between \nuc{182}{Pb} and \nuc{194}{Pb} at excitation energies
below 1~MeV. In particular, \nuc{186}{Pb} \cite{And00} is a unique example
of a fermion system where the two lowest  excited states 
are $0^+$ levels with  energies below 700 keV. The collective 
rotational bands built on top of these isomeric states suggest that 
they are associated with sizable deformations.

The ground state of Pb isotopes is known to be spherical down to 
\nuc{182}{Pb} \cite{Wau94}. The excited $0^+$ states have been interpreted
within two different frameworks, the mean-field and the shell models.
The apparently different mechanisms leading to these states 
within the two approaches have been shown to be complementary views 
of the same phenomenon \cite{Woo92}. Both models can explain the 
experimental data qualitatively. 

In a mean-field approach, the spectra of the Pb isotopes are understood 
as reflecting several competing minima in an axial quadrupole energy 
landscape, corresponding to spherical, oblate and prolate 
deformations. The very name ``shape coexistence phenomenon'' 
comes from this mean-field description \cite{Woo92,Hey83}. 
It is present in most nuclei around the Pb isotopes~\cite{Jul01} and
has been found also in other regions of the mass table~\cite{Woo92}. 
First calculations based on phenomenological mean fields 
have predicted the existence of several competing minima in the deformation
energy surface of neutron-deficient Pb isotopes \cite{May77,Ben89,Sat91} 
and a transition from oblate to prolate isomers \cite{Naz93}. 
More recently, several variants of the self-consistent mean-field
approach have confirmed these results \cite{Taj93,Lib99,Cha01,Nik02}.
However, in the neutron-deficient Pb region, shape coexistence 
cannot be completely described at the level of mean-field models.
The minima obtained as a function of the quadrupole moment
are often rather shallow and it is not clear \emph{a priori} whether they
will survive dynamical effects such as quadrupole vibrations.

In a shell model picture, the excited $0^+$ states are generated by 
$n$p-$n$h proton excitations across the \mbox{$Z=82$} shell gap, from the 
$3s_{1/2}$ level to the $1h_{9/2}$. The excitation energies are lowered 
by a residual quadrupole-quadrupole interaction. From this point of view,
the mean-field oblate minimum is associated with a proton 2p-2h 
configuration and the prolate one with proton 4p-4h intruder states 
\cite{Hey91}. Many-particle many-hole excitations cannot be easily
handled in full-scale shell-model calculations, in particular for
the large model space required for the description of heavy open-shell
nuclei. They are, therefore, treated with the help of algebraic models
\cite{Cos00,Fos03}.

A third, purely phenomenological, approach has also been used to interpret
the experimental findings: the 
shape-mixing picture \cite{Dup84}. In this model, the physically observed 
states are the result of the interaction between several configurations. 
They are thus a superposition of spherical, oblate and prolate configurations,
the relative weights in 
the mixing being determined by a fit to the experimental data.
Systematic two-level mixing calculations for \nuc{190-200}{Pb} have been
presented in Ref.\ \cite{Dup90}. A simple two-level mixing model has been 
applied to the analysis of the $\alpha$-decay hindrance factors 
in Ref.\ \cite{Del96}. Three level mixing has been performed 
in Ref.\ \cite{All98} for $^{188}$Pb. Let us also recall that, in such
models, the strength of monopole transitions is related to the value
of the interaction matrix elements between the unperturbed 
configurations \cite{Hey88}.

The aim of this paper is to provide a unified view of mean-field and 
shape mixing approaches. The method that we use has been presented 
in Ref.\ \cite{Val00} and applied to the study of shape coexistence 
in $^{16}$O in Ref.\ \cite{Ben03}. Results for $^{186}$Pb have already 
been published in Ref.\ \cite{Dug03a}. As a starting point, the 
method uses self-consistent mean-field wave functions generated as a 
function of the axial quadrupole moment. Particle number and angular 
momentum are restored by projecting these wave functions on the correct 
numbers of neutrons and protons and on spin. Finally, a mixing of the 
projected wave functions corresponding to different quadrupole moments is 
performed with a discretized version of the generator coordinate method. 
One of the appealing features of our method is that its only phenomenological 
ingredient is the effective nucleon-nucleon interaction which has 
been adjusted once and for all on generic nuclear properties. Another
attractive aspect is the direct determination of 
electric transition probabilities between any pair of states in the laboratory 
frame.

There have already been studies of the Pb isotopes along similar lines. 
Tajima \etal\ \cite{Taj93} and Chasman \etal\ \cite{Cha01} have used 
a very similar framework, but without any symmetry restoration. 
Mixing of microscopic self-consistent mean-field wave functions 
has been approximated  by Libert \etal\ \cite{Lib99} and Fleischer \etal\
\cite{Fle03} using a macroscopic Bohr Hamiltonian. Let us also mention 
that Fossion \etal\ \cite{Fos03} have performed 
three-configuration mixing calculations within the IBM formalism.
%
%
\section{Framework}
The starting point of our method is a set of HF+BCS wave functions
$| q \rangle$ generated by self-consistent mean-field calculations
with a constraint on a collective coordinate $q$. In the language of 
the nuclear shell model, such mean-field states incorporate 
particle-particle (pairing) correlations as well as many-particle-many-hole 
correlations by allowing for the deformation of the nucleus in the
intrinsic frame. As a consequence, however, such mean-field
states break several symmetries of the exact many-body states.
The symmetry violation causes some difficulties, for instance when 
relating the mean-field results to spectroscopic data obtained in 
the laboratory frame. Eigenstates of the angular momentum and the particle 
number operators are obtained by restoration of rotational and 
particle-number symmetry on each intrinsic wave function $| q \rangle$:
\begin{equation}
\label{eq:proj}
|J M q \rangle
= \frac{1}{{\mathcal N}_{JMq}}
  \sum_{K} g^{J}_{K}
  \hat{P}^J_{MK} \hat{P}_Z \hat{P}_N | q \rangle
,
\end{equation}
where ${\mathcal N}_{JMq}$ is a normalization factor;
$\hat{P}^{J}_{MK}$, $\hat{P}_N$, $\hat{P}_Z$ are projectors
onto the angular momentum $J$ with projection $M$ along the
laboratory z-axis and $K$ in the intrinsic z-axis, 
neutron number $N$ and proton number $Z$, respectively.
Here, we impose axial symmetry and time-reversal invariance
on the intrinsic states $| q \rangle$. Therefore, $K$ can only be
0 and one can omit the coefficients \mbox{$g^{J}_{K} = \delta_{K0}$}
as well as the sum over $K$. A variational configuration mixing on the 
collective variable $q$ is performed for each $J$-value separately:
\begin{equation}
\label{eq:discsum}
| J M k \rangle
= \sum_{q} f_{J,k} (q) | JM q \rangle
.
\end{equation}
The weight functions $f_{J,k}(q)$ are determined by requiring
that the expectation value of the energy
\begin{equation}
\label{eq:egcm}
E_{J,k}
= \frac{\langle JM k | \hat H | JM k \rangle}
       {\langle JM k | JM k\rangle} ,
\end{equation}
is stationary with respect to an arbitrary variation $\delta f^*_{J,k}(q)$.
This prescription leads to the discretized Hill-Wheeler equation
\cite{Hil53}. Collective wave functions in the basis of
the intrinsic states are then obtained from the set of weight
functions $f_{J,k}(q)$ by a basis transformation \cite{Taj93}.
In the $| J M k \rangle$ wave functions, 
the weight of each mean-field state $| q \rangle$
is given by:
\begin{equation}
\label{eq:weight}
g_{J,k} (q)
= \langle J M k | J M q \rangle
.
\end{equation}
Since the collective states $| JM k \rangle$ have good angular
momentum, their quadrupole moments and transition probabilities can
be determined directly in the laboratory frame without
further approximation. As the full model space of occupied 
single-particle states is used, no effective charge needs to be 
introduced. The formulae used to evaluate the overlap and Hamiltonian 
matrix elements have been presented in Ref.\ \cite{Val00}.

The same effective interaction is used to generate the mean-field wave
functions and to carry out the configuration mixing calculations. 
We have chosen
the Skyrme interaction SLy6 in the mean-field channel \cite{Cha98}
and a density-dependent zero-range pairing force with a strength of 
$-1250$~MeV fm$^3$ for neutrons and protons in connection with a soft 
cutoff at 5~MeV above and below the Fermi energy as defined 
in Ref.\ \cite{Rig99}.
%
%
\begin{figure}[t!]
\centerline{\epsfig{file=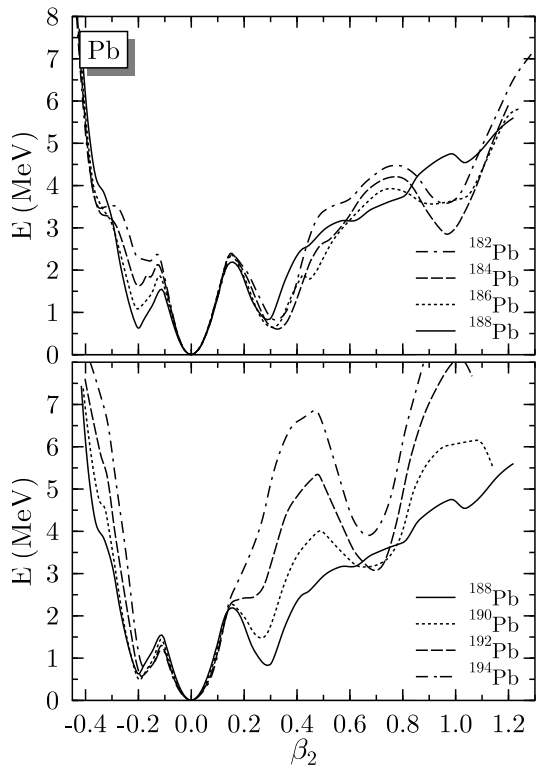}}
\caption{\label{fig:all:enz}
Particle-number projected deformation energy curves for \nuc{182-188}{Pb}
(upper panel) and \nuc{188-194}{Pb} (lower panel). The curve for 
\nuc{188}{Pb} is shown in both panels.
}
\end{figure}
%
%
A model that combines these extensions of the mean-field approach represents
a powerful tool. From a numerical point of view, it is still simple 
enough to be applied up to superheavy nuclei using the full model space 
of single-particle states with the proper coupling to the continuum.
Correlations corresponding to collective modes can be incorporated
step by step into the modeling and this helps to identify the relevant
degrees of freedom. The method has the advantage that results can
be interpreted within the intuitive picture of intrinsic shapes
and shells of single-particle states that is usually offered by 
mean-field models.

The model has already been successfully applied to the description
of coexisting structures in \nuc{16}{O} \cite{Ben03} and \nuc{32}{S}, 
\nuc{36,8}{Ar}, and \nuc{40}{Ca} \cite{BFH03}. Results for 
\nuc{186}{Pb} obtained with this model have already been presented 
in Ref.\ \cite{Dug03a}.
%
%
\section{Results}
\subsection{Potential landscapes}
The potential energy and other properties are plotted as a function 
of a dimensionless quadrupole deformation parameter approximately
independent of the nuclear mass:  
\begin{equation}
\label{beta2}
\beta_2
= \sqrt{\frac{5}{16 \pi}} \, \frac{4 \pi Q_{20}}{3 R^2 A}
,
\end{equation}
where $Q_{20}$ is the expectation value of the operator
\begin{equation}
Q_{\lambda,\mu}
= r^{\lambda} \; Y_{\lambda,\mu} 
\end{equation}
for $\lambda =2$ and $\mu =0$.

The particle-number projected potential energy curves  are displayed 
in Fig.\ \ref{fig:all:enz}. For all neutron numbers, the ground state 
is found to be spherical with a similar curvature of the energy curve 
around the spherical point. There is a well defined, slightly
deformed, oblate minimum for all isotopes. Its excitation energy 
increases while the depth of the well decreases when going down in
neutron number from \nuc{188}{Pb} to \nuc{182}{Pb}.
On the prolate side, an inflexion point in the \nuc{192-194}{Pb} curves
becomes a well-deformed minimum for \mbox{$\beta_2$} varying from 0.30 
for \nuc{190}{Pb} to 0.35 for \nuc{182}{Pb}. The deformation energy 
curves also present a deep minimum at superdeformed shapes for
\nuc{194-192}{Pb} which becomes shallower in \nuc{190}{Pb}, and disappears 
for the lighter isotopes. For the three lightest isotopes,  there is an
additional minimum at still larger deformations, with a maximal depth of
1.5~MeV for \nuc{184}{Pb}. 
%
%
\begin{figure}[t!]
\centerline{\epsfig{file=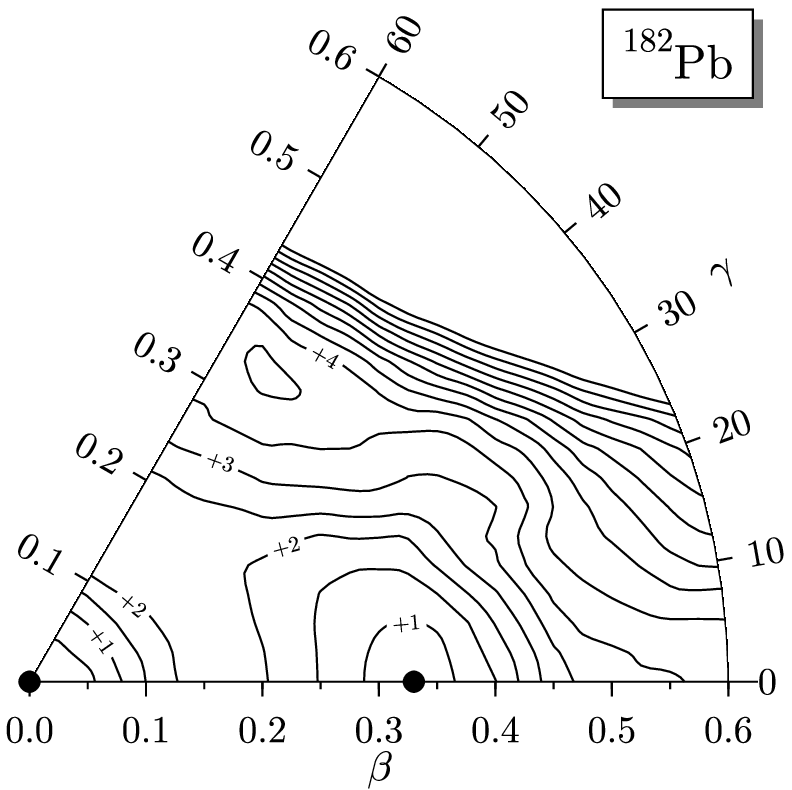,width=5.0cm}}
\centerline{\epsfig{file=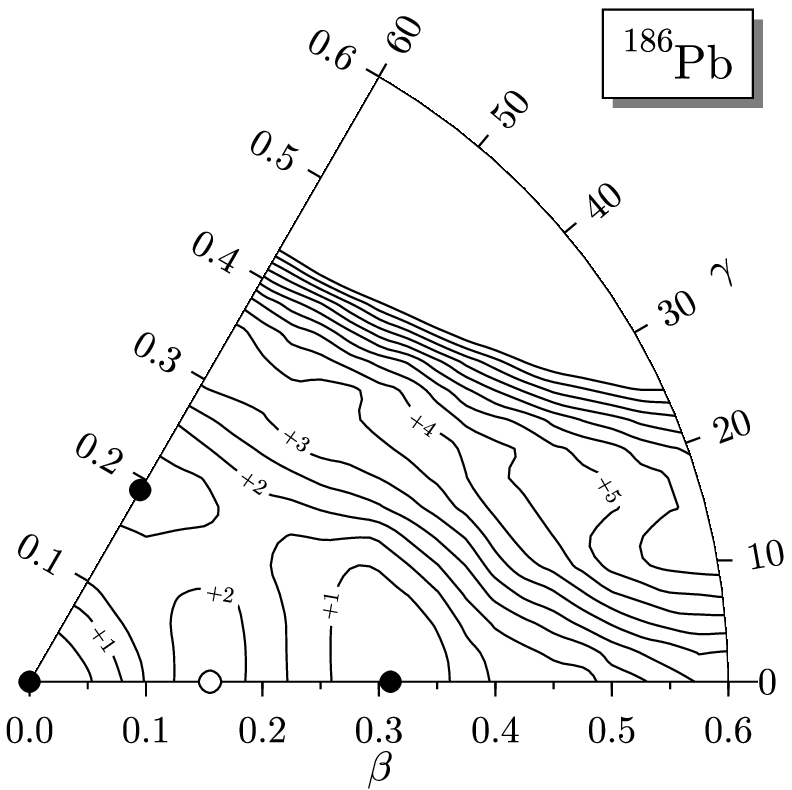,width=5.0cm}}
\centerline{\epsfig{file=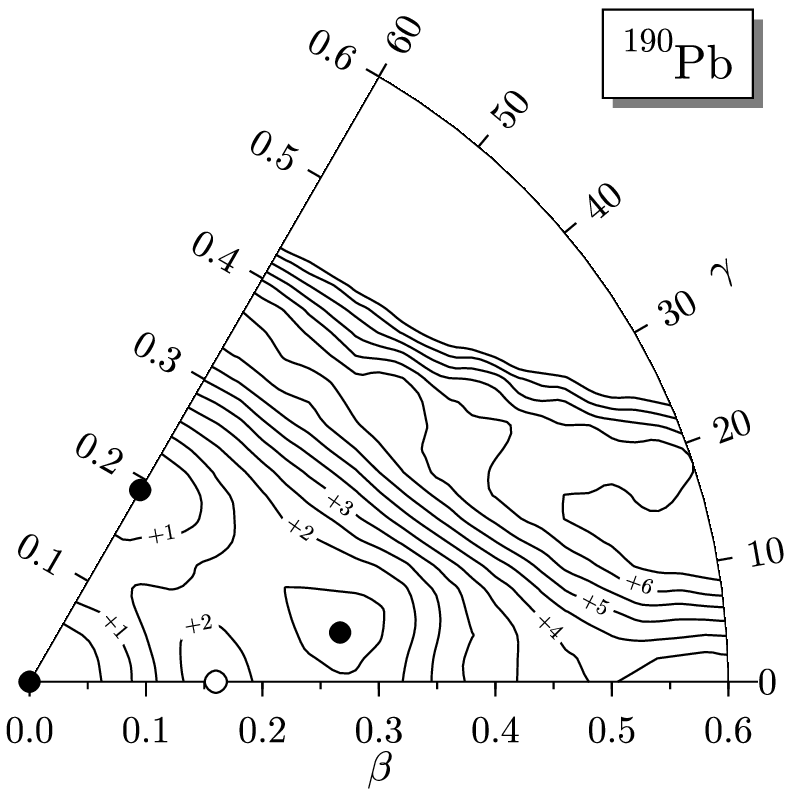,width=5.0cm}}
\centerline{\epsfig{file=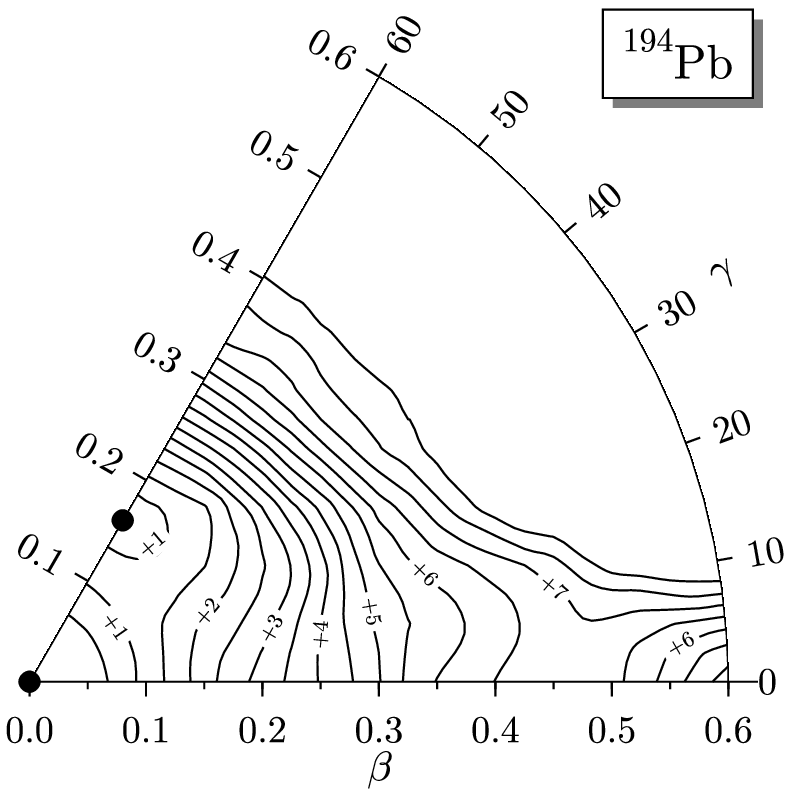,width=5.0cm}}
\caption{\label{fig:all:betagamma:enz}
Particle-number projected deformation energy surfaces for the Pb
isotopes as indicated. The contour lines are separated by 0.5~MeV.
Filled circles denote minima, open circles maxima of the potential 
landscape.}
\end{figure}
%
%

Exploratory studies performed by S.~{\'C}wiok \cite{Cwi00}
indicate that calculations performed with a zero-range volume 
pairing interaction with appropriate strengths result in
oblate and prolate minima at energies close to the experimental 
values. Comparing our results with those presented in Ref.\  
\cite{Fle03}, one also has to conclude that the form factor chosen 
for the pairing interaction has some influence on the relative 
position of the minima in \nuc{186}{Pb}. However, since our aim 
is to determine the influence of correlations 
beyond mean-field  for interactions that have been validated 
on a large set of data, we will limit ourselves to the same 
form of pairing as used in previous works.

For deformations between $\beta_2=-0.3$ to $+0.4$, the variation of
energy does not exceed 2.5~MeV in \nuc{182-190}{Pb}. The barrier 
heights between the minima are slightly larger than 2~MeV  for 
the prolate and  some superdeformed minima. 
In contrast, the barrier heights between  spherical 
and oblate minima is always small. With such a topology,
most minima of the potential energy curves cannot be unambiguously
identified with physical states and one can speculate whether these 
structures may be washed out by the vibrational fluctuations 
associated with quadrupole motion. As discussed on the basis of 
phenomenological energy maps for neutron-deficient Pb isotopes by
Bengtsson and Nazarewicz \cite{Ben89}, it is hard to infer the effect
of vibrational fluctuations only from the topology of the potential 
landscape.
%
%
\begin{figure}[t!]
\centerline{\epsfig{file=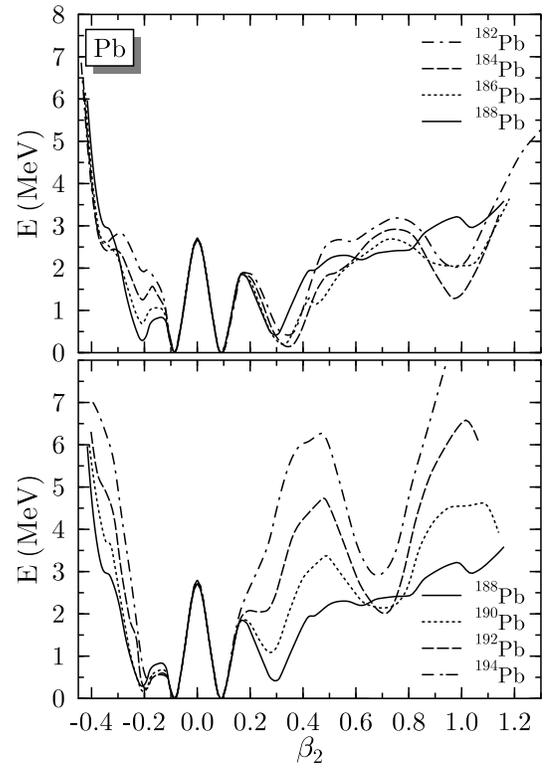}}
\caption{\label{fig:all:ejnz}
Particle-number and angular momentum \mbox{$J=0$} projected deformation 
energy curves for \nuc{182-194}{Pb} drawn in the same manner as in 
Fig.\ \protect\ref{fig:all:enz}. All curves are normalized to the
energy minimum.
}
\end{figure}
%
%

In principle, oblate and prolate minima found by a calculation 
limited to axial deformation
might be connected by a path through triaxial quadrupole deformation.
To verify that this is not the case for the Pb isotopes, we have
calculated the two-dimensional energy maps for four isotopes.
They are displayed in Fig.\ \ref{fig:all:betagamma:enz}. 
The triaxial deformation angle $\gamma$ is defined by:
\begin{equation}
\tan (\gamma) 
= \sqrt{2} \; \frac{Q_{22}}{Q_{20}} 
.
\end{equation}
The potential maps for \nuc{182}{Pb} and \nuc{194}{Pb} exhibit 
two minima at small deformations, spherical and oblate for 
\nuc{194}{Pb} and spherical and prolate for \nuc{182}{Pb},
while three stable minima can be seen for the two other isotopes.
The barrier height between the prolate and oblate minima is larger 
than 500 keV. In \nuc{190}{Pb}, the prolate minimum is slightly 
triaxial with \mbox{$\gamma \approx 10^o$}.

The energy curves obtained after projection on angular momentum 
\mbox{$J=0$} are shown in Fig.\ \ref{fig:all:ejnz}. 
As can be seen, the conclusions drawn from 
Fig.\ \ref{fig:all:enz} remain qualitatively valid. The spherical 
mean-field state is rotationally invariant and, therefore, contributes 
to \mbox{$J=0$} only. As found in previous studies of spherical 
nuclei with our method, there are two minima at small deformations. 
They do not represent distinct physical states, 
but, as will be clear after the configuration-mixing calculation, 
the correlated spherical state. The depth of the oblate well is 
affected by angular momentum projection, as the barrier separating it 
from the spherical minimum is quite low for all neutron numbers.
%
%
\subsection{Single-particle spectrum}

%
%
\begin{figure}[t!]
\centerline{\epsfig{file=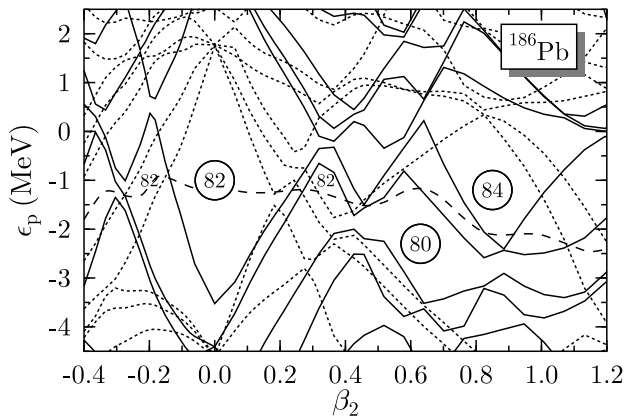}}
\caption{\label{fig:pb186:sp:p}
Proton eigenvalues of the single-particle Hamiltonian for \nuc{186}{Pb}.
Solid lines denote states with positive parity, dotted lines 
states with negative parity. The Fermi energy is plotted with
a dashed line.
}
\end{figure}
\begin{figure}[t!]
\centerline{\epsfig{file=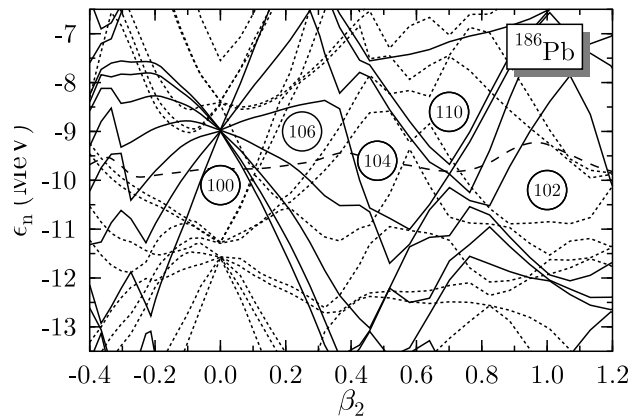}}
\caption{\label{fig:pb186:sp:n}
Eigenvalues of the single-particle Hamiltonian obtained for 
neutrons in \nuc{186}{Pb}.
Solid lines denote states with positive parity, dotted lines 
states with negative parity. The Fermi energy is plotted with
a dashed line.
}
\end{figure}

%
%
The minima obtained for specific deformations are associated with
a shell effect due to a
low level density around the Fermi energies for both protons and
neutrons. To visualize this effect, we show Nilsson plots of the
proton and neutron single-particle energies for \nuc{186}{Pb} in 
Fig.~\ref{fig:pb186:sp:p} and Fig.~\ref{fig:pb186:sp:n},
respectively. Besides an overall offset due to the 
change in proton-neutron asymmetry, the proton spectra are 
similar for other neutron numbers. These curves are also quite close 
to those presented in Ref.\ \cite{Taj93}, where
a different set of Skyrme and pairing parameters was used. 

The spherical gap (\mbox{$\beta_2 = 0$}) dominates the proton spectrum,
There is also a spherical neutron subshell closure at \mbox{$N=100$} 
between the $2f_{7/2^-}$ and $1i_{13/2^+}$ levels.
Several additional, smaller gaps are visible at various deformations;
let us first focus on the oblate side.

A large proton shell gap is present for an oblate deformation 
$\beta_2 \approx -0.2$. Comparing the level ordering at 
this deformation with that at sphericity indicates that the 
$3s_{1/2^+}$ levels have been pushed up from below the Fermi energy and 
have crossed two down-sloping $1h_{9/2^-}$ levels. The deformation 
at which the first level crossing occurs depends predominantly on 
the size of the spherical \mbox{$Z=82$} gap: the larger the gap, 
the larger the deformation of the oblate minimum, as this crossing 
corresponds to the position of the barrier between the
oblate and the spherical energy minima. In a shell
model picture~\cite{Hey91,Cos00}, this oblate state is described 
by a 2p-2h excitation from the $3s_{1/2^+}$ level to the 
$1h_{9/2^-}$ one.

The size of the oblate proton gap is mainly determined by the 
splitting of the $1h_{9/2^-}$ state with deformation, which can 
be expected to be force independent. It depends also on the 
energy difference between the $1h_{9/2^-}$ and the $1i_{13/2^+}$ 
levels, which has to be large enough so that the $1i_{13/2^+}$ states 
diving down with oblate deformation stay above the oblate 2p-2h gap.
That the level closest to the Fermi level is a $1h_{9/2^-}$ state
is supported by the fact that in adjacent even-odd Tl and Bi nuclei, 
states built on the proton $1h_{9/2^-}$ orbital above and the $3s_{1/2^+}$ 
orbital below the \mbox{$Z=82$} gap have been observed at excitation 
energies of only a few hundred keV \cite{Lan95}.

This interpretation of the oblate states is also consistent with the 
observation of $11^-$ high-spin isomers for \nuc{188-198}{Pb}\cite{Jul01}. 
These isomers are interpreted as broken-pair 
proton $(3s_{1/2}^{-2}1h_{9/2}1i_{13/2})_{11^-}$ states. The values
of their $g$ factor and of their spectroscopic quadrupole moment
have been measured in \nuc{194-196}{Pb}~\cite{Vyv02}
and are consistent with excitations built on top of the excited $0^+$ 
oblate states. They have been successfully described in this
way by Smirnova \etal\ \cite{Smi03a}. 

The level ordering shown in Fig.~\ref{fig:pb186:sp:p}
is thus corroborated by several
experimental data and gives a coherent picture of the oblate minimum.
The situation is less evident for the neutrons.
As can be seen from Fig.~\ref{fig:pb186:sp:n}, the levels
which are responsible for the oblate neutron  gap 
originate from the $1h_{9/2^-}$ and $1i_{13/2^+}$ shells. Their first
crossings occur 
at deformations around $\beta_2 \approx -0.2$; only two
$1i_{13/2^+}$ levels have energies lower than the Fermi energy
at the oblate minimum of  \nuc{186}{Pb}.
%
%
%
%
%
As we shall show in the next sections, the energy of the predominantly 
oblate $0^+$ state is overestimated for all isotopes by our model. 
A possible cure to this deficiency could thus be to slightly decrease 
the energy of the $1i_{13/2^+}$ neutron level. 
A larger occupation of these neutron states with large $m_z$ 
values would increase the quadrupole interaction between neutrons 
and protons, the proton levels close to the Fermi energy having
also large $m_z$ values. This would decrease the energy of the
oblate configuration and modify its dependence on the neutron number.
There are other evidences that the $1i_{13/2^+}$ energy is 
overestimated. On the prolate side of deformations, levels from 
the spherical proton $1i_{13/2^+}$ and $1h_{9/2^-}$ shells 
come close to the Fermi energy in the ground state of 
transactinides. An analysis of the quasiparticle spectrum of 
\nuc{249}{Bk} performed in the cranked HFB approach using the Skyrme 
interaction SLy4 in combination with the same pairing 
prescription used here indicates that the $1i_{13/2^+}$ state is
predicted too high above the $1h_{9/2^-}$ orbitaland should be lowered
\cite{BBDH03}. A similar conclusion emerges from the analysis of the 
single-particle spectra in \nuc{208}{Pb}, where most mean-field
models predict the proton $1i_{13/2^+}$ shell too far above 
the $1h_{9/2^-}$ level \cite{Ben99,Kle02}. As we discussed above, 
however,  these single-particle energies are also constrained by the
properties of oblate states around \nuc{186}{Pb} and 
the distance between these levels may not be changed by 
more than a few hundreds of keV.

%
\begin{figure*}[t!]
\centerline{\epsfig{file=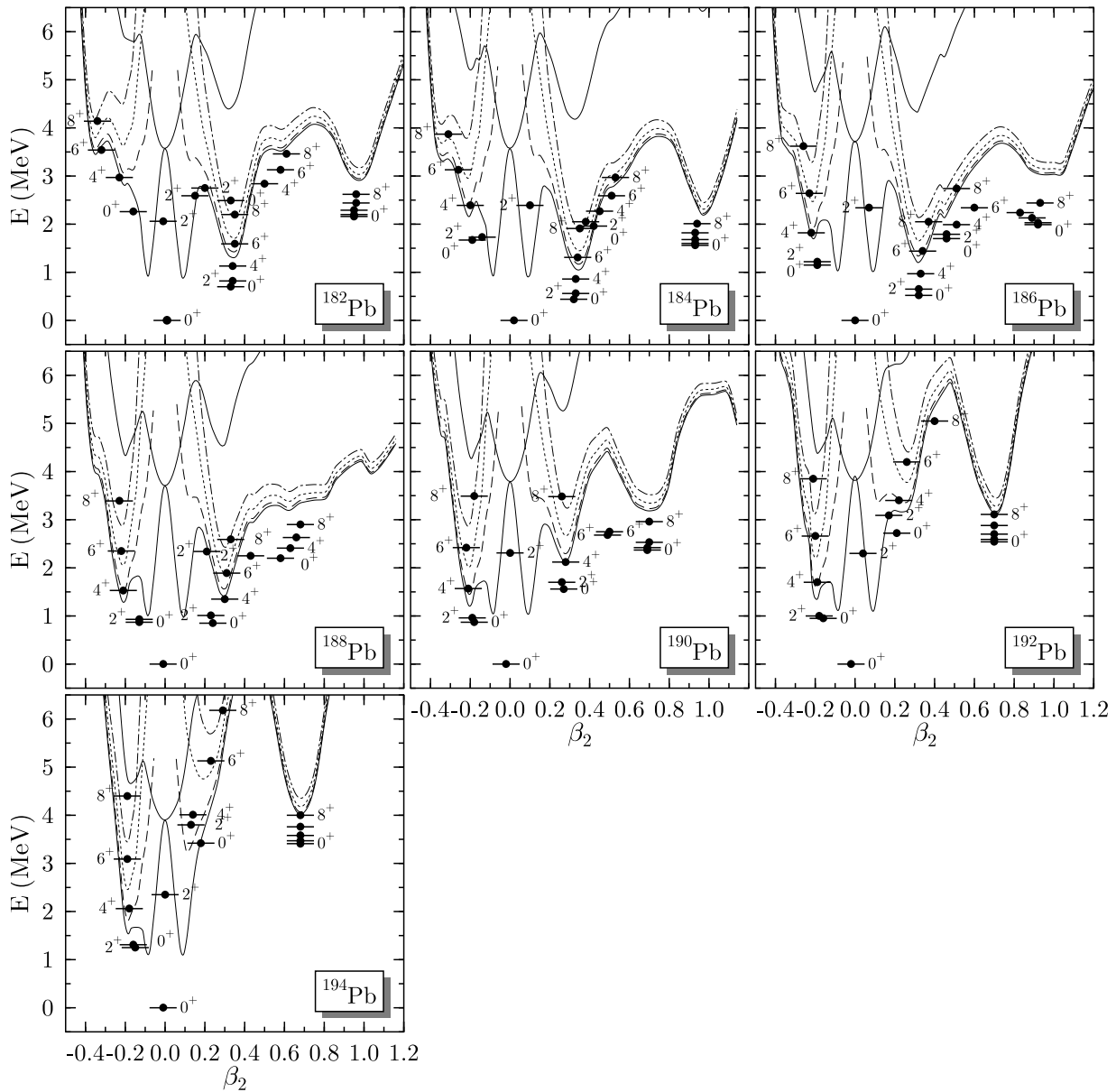}}
\caption{\label{fig:all:ex1}
Excited states in \nuc{182-194}{Pb}. Only the lowest band structures 
are shown. For the superdformed bands, only the $0^+$ and $8^+$ 
levels are labeled.
}
\end{figure*}
%
%

The prolate minimum is not 
as nicely related to a deformed proton shell effect, as no 
large gap is obtained for a deformation around $\beta_2$ equal 
to 0.3. The relation of this prolate configuration to a 4p-4h shell 
model configuration is therefore not evident. In contrast, there is 
a small gap at the Fermi surface for \mbox{$\beta_2$} values around
0.35, which corresponds to a 6p-6h configuration. A similar result
has been obtained with the potential models used for 
microscopic-macroscopic calculations of Ref.~\cite{Ben89}.

As can be seen from Fig.~\ref{fig:pb186:sp:n}, there are sizable 
neutron gaps at prolate deformations, which can be associated with 
structures in the potential energy curves, in particular
at \mbox{$N=102$}, \mbox{$N=104$} and \mbox{$N=106$} for  
\mbox{$\beta_2 \approx 0.4$}. These can be expected to enhance the
depth of the prolate minimum in the potential landscape. Above 
\mbox{$N=106$}, the Fermi energy crosses a region of high level 
density, which explains the disappearance of the prolate minimum
at neutron numbers above \nuc{190}{Pb}. The \mbox{$N=102$} and
\mbox{$N=110$} gaps at larger deformations can be associated with the 
superdeformed minima in \nuc{184}{Pb} and \nuc{192}{Pb} respectively.
%
%
\subsection{Excitation spectra}
\begin{figure}[t!]
\centerline{\epsfig{file=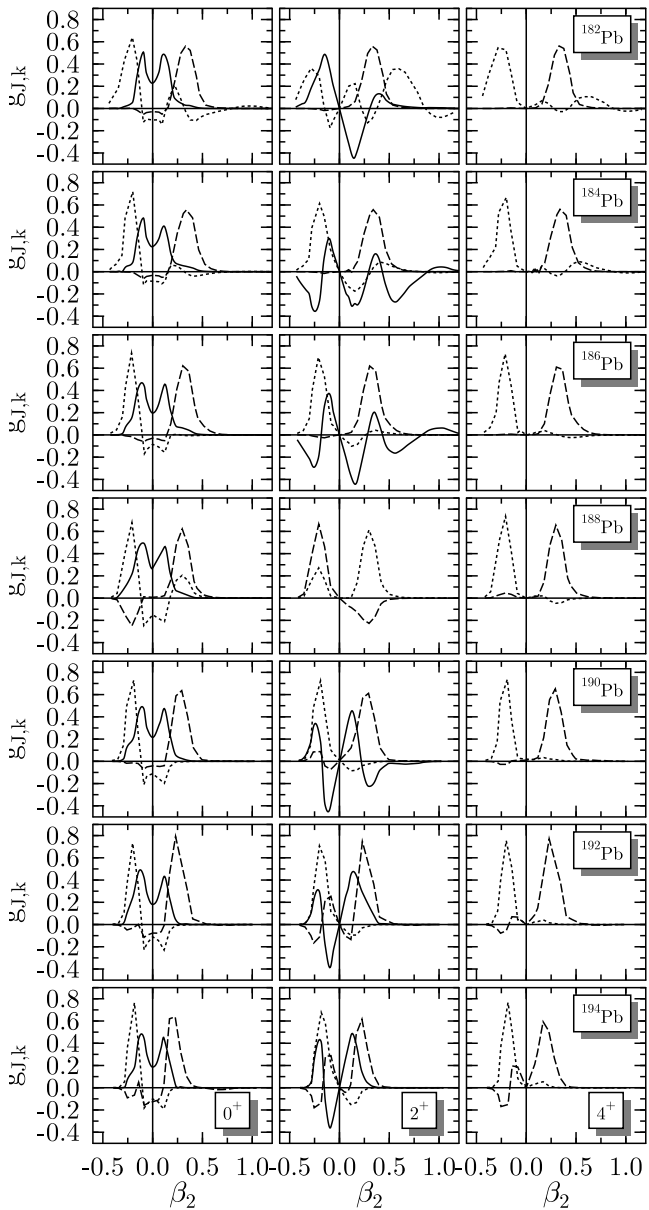}}
\caption{\label{fig:all:wf1}
Collective wave functions of spherical (solid lines), oblate (dotted lines)
and prolate (dashed lines) states for \nuc{182-194}{Pb}. 
}
\end{figure}

The excitation spectra of \nuc{182-194}{Pb} obtained after configuration
mixing are presented in Fig.\ \ref{fig:all:ex1}. 
The bars representing each state are plotted at a mean deformation 
$\bar{\beta}_{J,k}$ in the intrinsic frame defined as:
\begin{equation}
\bar{\beta}_{J,k}
= \int \! d\beta_2 \; \beta_2 \; g_{J,k}^2 (\beta_2)
,
\end{equation}
where $\beta_2$ is related to the value of the constraint used 
to generate the mean-field states (see eq.~(\ref{beta2})). 
This average value does not correspond to any 
observable, but is convenient to characterize the decomposition 
of each collective state into its mean-field components The collective 
wave functions for selected states are shown in Fig.\ \ref{fig:all:wf1}. 
The fully correlated ground state is dominated by mean-field
configurations close to sphericity for all isotopes, in agreement 
with the data. In \nuc{190-194}{Pb}, the 
first excited $0^+$ levels are dominated by oblate mean-field configurations.
In \nuc{188}{Pb}, a configuration dominated by prolate mean-field states
is nearly degenerate with the oblate one; for lighter isotopes, 
it becomes the first excited state. 
%
%
%
%

For several isotopes, a classification of most states into rotational 
bands emerges quite naturally from the value of the average deformation 
$\bar{\beta}_{J,k}$. The simplest case is \nuc{190}{Pb}, where rotational 
bands above the oblate, prolate and superdeformed configurations are 
easily identified in Fig.\ \ref{fig:all:ex1}. An exception are the two 
nearly-degenerated $6^+$ levels found at intermediate deformations
between the prolate and superdeformed bands which are very close in energy
and, thus, strongly mixed. There is also an isolated 
$2^+$ state with $\bar\beta_{2,3}$ close to zero at 2.3~MeV excitation
energy. Such a state is present at a similar energy in all other
Pb isotopes studied here. The situation in \nuc{186}{Pb} is nearly 
as simple, except for two additional bands with average 
deformations between those of the first prolate and the superdeformed states. 
In the two lighter isotopes, the mean weight of oblate states in the
predominantly oblate band is moving toward larger deformations
with increasing spin, while the prolate band has a much more stable 
mean deformation. For \nuc{192}{Pb} and \nuc{194}{Pb} there are
many levels above 4~MeV excitation energy that cannot be easily 
associated with a rotational band. Figure~\ref{fig:all:wf1} 
indicate that the collective wave functions 
$g_{J,k}$ are spread over a large range of 
deformed configurations. For \mbox{$J=0$} and 2, this spreading 
extends on both the prolate and oblate sides, while for \mbox{$J=4$},
it is more limited to the vicinity of either an oblate or a prolate 
configuration. Unfortunately, it is difficult to represent this spreading 
by the mixing of only two or three configurations, as done in 
schematic models. A reduction of the configuration mixing basis to 
an oblate, prolate and spherical state leads to much too small a
configuration mixing. Of course, the classification in bands that 
we have obtained from the $\bar\beta_{J,k}$ valueshas to be confirmed by 
the behavior of the calculated transition probabilities between 
the various states.
%
%
\begin{figure}[t!]
\centerline{\epsfig{file=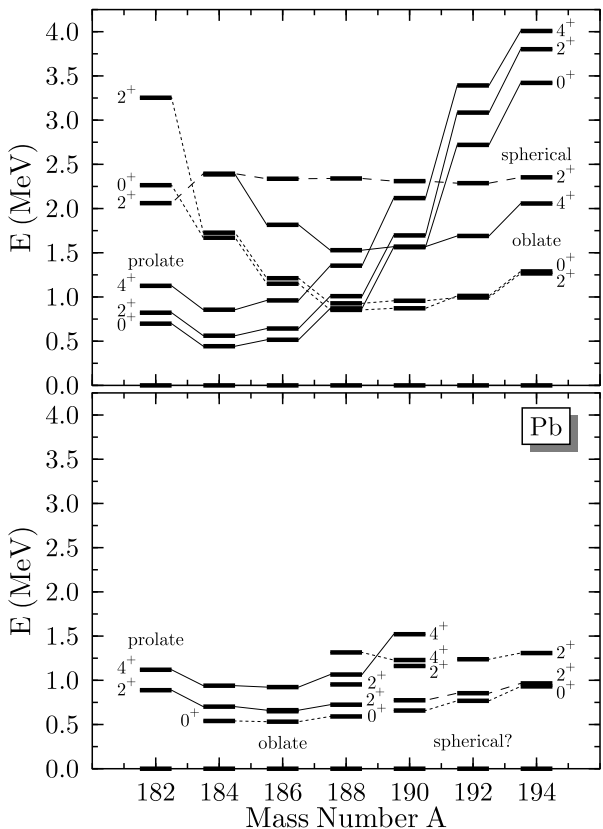}}
\caption{\label{fig:all:bands}
Systematics of calculated (upper panel) and experimental (lower panel) 
excited states in \nuc{182-194}{Pb}.
}
\end{figure}
%
%

Our results concerning the lowest bands found in all isotopes are 
summarized in Fig.\ \ref{fig:all:bands} (upper part) and compared to 
experimental data (lower part). Both for experiment and theory, the 
bands are sorted as oblate, prolate or spherical on the basis of 
their spectroscopic properties.

Experimental trends are qualitatively reproduced by our calculation,
although changes with neutron number are too abrupt. Our calculation 
predicts nearly degenerate oblate and prolate $0^+$ excited states
in \nuc{188}{Pb} only, while experimentally the two first excited 
$0^+$ states lie within 100 keV in \nuc{184-188}{Pb}.
A systematic discrepancy between theory and experiment is the 
position of the $2^+$ level interpreted as a vibration of the 
spherical ground state. Its energy is larger than 2~MeV in our 
calculation for all isotopes, a value 
to be compared with an experimental energy around 1~MeV \cite{Dra03}.
Looking back into Fig.~\ref{fig:all:wf1}, a plausible reason for the better
agreement obtained for the prolate states than for the other levels may lie
in the behavior of the wave functions. For all isotopes, the ground
state $0^+$ wave functions have an extension on the oblate side 
similar to the predominantly oblate wave function. On the other hand, 
the amplitude of this wave function is negligible for deformations 
close to that of the maximum of the prolate wave function. Therefore,
the configurations close to sphericity should be expected to be coupled 
more strongly to the oblate configurations than to the prolate ones. 
Such a strong coupling pushes the predominantly oblate $0^+$ states 
higher in energy than the predominantly prolate ones. A similar 
explanation can be invoked to account for the too high energy of the 
vibrational $2^+$ level constructed on the ground state. The wave 
function of this level extends on both prolate and oblate sides up to the 
tails of the oblate and prolate $2^+$ wave functions. A possible way to
reduce these large overlapping regions of collective wave functions could
be the inclusion of triaxial deformation, which might lead
to a different spreading of wave functions in the $\gamma$ direction. 

Experimental prolate states are nicely described by our calculations, 
although the increase of their excitation energies with mass is 
slightly too fast for the heaviest isotopes. The simple picture 
of shape isomerism, as suggested by the potential landscapes,
is supported by our calculations for oblate states, even after the
inclusion of dynamical correlations on top of the shallow minima. This 
result is by far non trivial. However, the position of these states 
is slightly too high in energy and the trend with neutron number is 
not correct. In particular, the excitation energy of the oblate isomer 
is not minimal at mid-shell as it is experimentally. The minimum is 
shifted by four mass units towards heavier systems. The same situation 
is encountered in calculations using the Gogny force \cite{Cha01}. 
%
%
\begin{figure}[t!]
\centerline{\epsfig{file=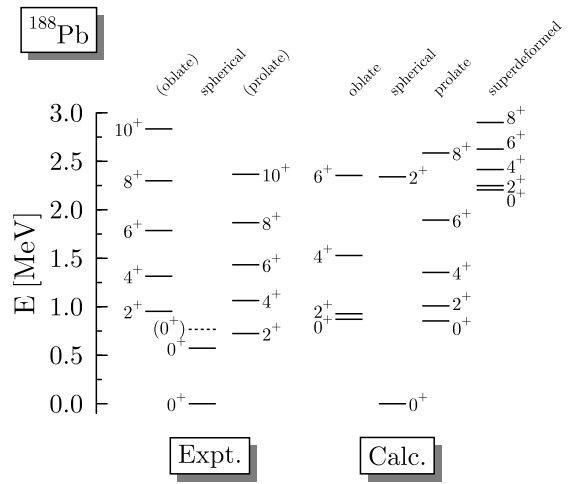}}
\caption{\label{fig:pb188:spect}
Comparison between the calculated excitation energies and the available
experimental data for low-lying states in \nuc{188}{Pb}. Data are
taken from \protect\cite{Dra03}.
}
\end{figure}
%
%

There are very recent new experimental data for \nuc{188}{Pb}
\cite{Dra03,Dew03}. In particular, this isotope is the one 
for which there are the most extensive data on $E0$ and $E2$ 
transition probabilities. For this reason, we show a direct 
comparison between theoretical and experimental spectra for 
this nucleus in Fig.~\ref{fig:pb188:spect}. The main discrepancies 
with the data are for spins larger than 4 where the theoretical 
spectra are too spread. Experimentally, there is no clear evidence 
that the two first excited $0^+$ levels are predominantly prolate or 
oblate in character. Starting at \mbox{$J=2$}, the states may be sorted 
in two bands fitted by a Harris expansion. This is also the case in 
our calculation. Experimentally, the lowest $2^+$ state is interpreted 
as a member of the prolate band, while it belongs to the oblate one 
in our calculation.
%
%
\subsection{Correlation energies}
From Fig.~\ref{fig:all:ex1}, one can extract 
the correlation energies due to the restoration of the rotational symmetry 
and the removal of axial quadrupole vibrations. Projection on particle
number is already included in the mean-field energy curves. The magnitude 
of the correlation energy due to this projection has been determined in 
a previous study of Pb isotopes \cite{Hee01}, and is around 1.0~MeV.
The ground states are found to be spherical, irrespective of the nature of 
the correlations that are introduced. The rotational energy has a value 
close to 2.5~MeV in all cases and vibrational correlations bring an 
additional energy gain of approximately 1.0~MeV. The total energy gain due 
to these two correlations is rather stable for \nuc{182-194}{Pb} 
and is comprised between 3.5 and 4.5~MeV.
%
%
\subsection{Quadrupole deformation}
%
%
\subsubsection{Spectroscopic quadrupole moment}
The spectroscopic quadrupole moment is given by:
\begin{eqnarray}
\label{eq:Qs}
Q_c (J_k)
& = & \sqrt{\frac{16\pi}{5}} \,
      \langle J, M=J, k | \hat{Q}_{2 0} | J, M=J, k \rangle
      \nn \\
& = & \sqrt{\frac{16\pi}{5}} \,
            \frac{\langle J J 2 0 | J J \rangle}{\sqrt{2J+1}}
      \nn \\
&   & \times
      \sum_{q,q'} f_{J,k}^{*} (q ) \,
                  f_{J,k}     (q') \,
            \langle J q || \hat{Q}_{2 0} || J q' \rangle ,
\end{eqnarray}
%
%
\begin{figure}[t!]
\centerline{\epsfig{file=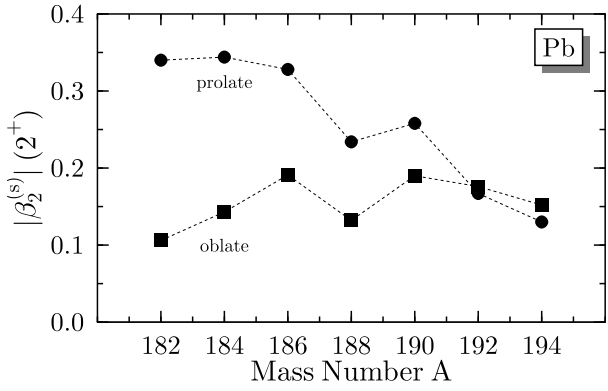}}
\caption{\label{fig:all:qspect}
Systematics of the intrinsic deformation $\beta^{(s)}_2$ derived from the
absolute value of the spectroscopic quadrupole moment of low-lying 
$2^+$ states in \nuc{182-194}{Pb}. Circles denote prolate, squares 
oblate levels.
}
\end{figure}
%
%
where $\hat{Q}_{2 0}$ is the proton quadrupole operator. Although directly
accessible in a model-independent way in the laboratory frame, 
the spectroscopic 
quadrupole moment has the disadvantage that its value scales with mass 
and angular momentum. It is then difficult to compare different 
nuclei or different members of the same rotational band.
Using the static rotor model, one can define a dimensionless quadrupole 
deformation in the intrinsic frame $\beta_2^{(s)}(J_k)$ which is easier 
to visualize and to relate to the deformation parameters introduced in 
most other models:
\begin{eqnarray}
\beta_2^{(s)}(J_k)
& = & \sqrt{\frac{5}{16 \pi}} \, \frac{4 \pi Q_2^{(s)}(J_k)}{3 R^2 Z}
      \nn \\
Q_2^{(s)}(J_k)
& = & - \frac{2J+3}{J} \, Q_c (J_k)
\end{eqnarray}
with $R=1.2 \, A^{1/3}$ and \mbox{$K=0$}.

The absolute value of the deformation parameter $\beta_2^{(s)}$ derived from 
the spectroscopic quadrupole moment of low-lying $2^+$ states is shown 
in Fig.~\ref{fig:all:qspect}. The $\beta_2^{s}$ values corresponding to 
higher-lying members of the prolate and oblate rotational bands are 
in most cases quite similar to those obtained for the $2^+$ state.
In agreement with the systematics of the deformed minima in the 
potential energy surfaces shown in Figs.~\ref{fig:all:enz} 
and~\ref{fig:all:ejnz}, the deformation of the prolate $2^+$ levels
increases with decreasing neutron number, while the deformation
of the oblate $2^+$ state stays fairly constant. The small values
of $|\beta_2^{(s)}|$ found for the oblate $2^+$ states in \nuc{182-184}{Pb} 
and the prolate $2^+$ states in \nuc{192-194}{Pb} can be related to 
an increased spreading of the corresponding wave function into the 
spherical well due to the very small or even vanishing potential barrier
in these cases. The discontinuity in the systematics of $\beta_2^{(s)}$
for oblate and prolate states predicted for \nuc{188}{Pb} is probably
related to the increased mixing of all low-lying states due to their 
near-degeneracy.
%
%
\subsubsection{Reduced transition probabilities}
%
%
\begin{figure}[t!]
\centerline{\epsfig{file=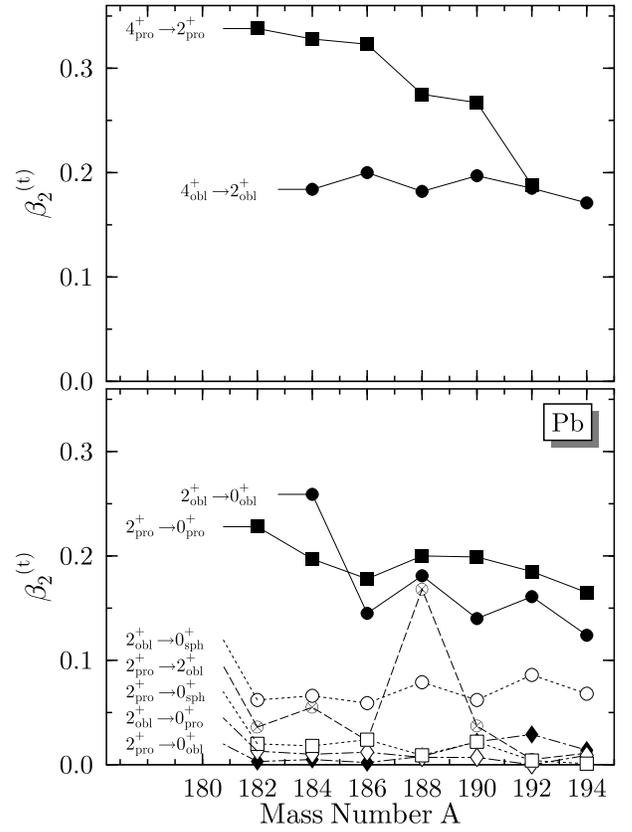}}
\caption{\label{fig:all:be2}
Systematics of the intrinsic deformation derived from the reduced 
$E2$ transition probability between low-lying $2^+$ and $0^+$ states 
(lower panel) and low-lying $4^+$ and $2^+$ states (upper panel)
in \nuc{182-194}{Pb}. For in-band transitions between higher-lying 
states, $\beta_2^{(t)}$ values are similar to those obtained for
the \mbox{$4^+ \to 2^+$} transitions.
}
\end{figure}
%
%
The reduced $E2$ transition probability is determined from
\begin{eqnarray}
\lefteqn{
B(E2; J_{k'}' \to J_k)}
      \nn \\
& = & \frac{e^2}{2J'+1}
      \sum_{M =-J }^{+J }
      \sum_{M'=-J'}^{+J'}
      \sum_{\mu=-2}^{+2}
      | \langle J M k | \hat{Q}_{2 \mu} | J' M' k' \rangle |^2
      \nn \\
& = & \frac{e^2}{2J'+1}
      \left| \sum_{q,q'} 
      f^{*}_{J,k} (q) \, f_{J',k'} (q') \,
      \langle J q || \hat{Q}_{2} || J' q' \rangle
      \right|^2
.
\end{eqnarray}
The $B(E2)$ values are often given in Weisskopf units (W.u.),
where 1~W.u.\ is equal to $5.940 \times 10^{-2} \; A^{4/3}$ $e^2$ fm$^4$.
Again, the $B(E2)$ value has the advantage that it can be deduced in a
model-independent way in the laboratory frame, but also has the disadvantage 
that it scales with mass and angular momentum. Using the 
static rotor model, the $B(E2)$ values are related to a dimensionless 
deformation in the intrinsic frame
\begin{equation}
\beta_2^{(t)}(J^\prime_{k'} \to J_k)
= \frac{4 \pi}{3 R^2 Z}
  \sqrt{
  \frac{B(E2; J^\prime_{k'} \to J_k)}
       {\langle J' \,0 \, 2 \, 0 | J \, 0 \rangle^2 e^2}}
\end{equation}
with \mbox{$R=1.2 \, A^{1/3}$}. This transition quadrupole moment reflects
the intrinsic deformation of the states if, and only if, both states
involved have a similar structure. Differences between
$\beta_2^{(t)}(J^\prime_{k'} \to J_k)$ and 
$\beta_2^{(s)}(J^\prime_{k'})$ give a measure of the validity of the
static rotor model for a given band.

The $\beta_2^{(t)}$ values derived from the transition probabilities 
between low-lying $4^+$, $2^+$ and $0^+$ states are presented in 
Fig.~\ref{fig:all:be2}. The values calculated for in-band transitions 
for higher $J$ values are close to those for the $4^+ \to 2^+$ ones. 
Except for \nuc{184}{Pb}, very similar $\beta_2^{(t)}$ values are
obtained for all transitions within the oblate band.  On the contrary,  
the deformation determined from the $2^+_{\rm pro} \to 0^+_{\rm pro}$ 
transition is significantly smaller than that calculated 
using the transition starting from the prolate $4^+$ state.
This is due to a change of structure of the collective wave functions
with spin: the  $0^+$ wave functions are much more mixed than 
the wave functions corresponding to higher $J$-values.
The particularly large $\beta_2^{(t)}$ value found for the 
$2^+_{\rm pro} \to 2^+_{\rm obl}$ transition in \nuc{188}{Pb} again
reflects the large mixing of these two nearly degenerate states.
From the right part of the figure, one sees that the in-band 
transitions are in most cases approximately one order of magnitude 
more intense than the out-of-band ones.
Comparing the deformations calculated from the spectroscopic
moments and shown in Fig.~\ref{fig:all:qspect} to those coming 
from the transition moments, Fig.~\ref{fig:all:be2}, one sees 
a close similarity, except for the oblate states
in the two lightest isotopes and in \nuc{188}{Pb}. In these 
three cases, this can be viewed as a confirmation that
the states labeled oblate result in fact from the mixing of a large
range of mean-field states and that these mixings vary with angular momentum.  
The existence of rather pure rotational bands is confirmed
for nuclei where both $\beta_2^{(t)}$ and $\beta_2^{(s)}$
have close values. 

To the best of our knowledge, there are no experimental data for 
$B(E2)$ values in neutron-deficient Pb isotopes, except the very 
recent measurements in \nuc{188}{Pb} of Dewald \etal\ \cite{Dew03}. 
The $B(E2)$ value measured for the $4_1^+ \rightarrow 2_1^+$  
transition is equal to 160~W.u.\ and that
for the $2_1^+ \rightarrow gs$ transition is 5.3~W.u.
These values are in between those that we obtain for
transitions starting from the prolate and oblate $4^+$ and $2^+$ states.
The calculated $B(E2)$ values
for the $4_1^+ \rightarrow 2_1^+$ transition is
equal to 288~W.u.\ for the prolate band and 126~W.u.\ for the oblate one,
while the out-of-band transitions to the ground state are 0.2~W.u.\ and 
17.0~W.u.\ respectively. This result probably reflects the fact
that the calculation underestimates 
the configuration mixing for the $2^+$ or the $4^+$ states.
Dewald \etal\ have determined a $\beta_2$ value
of 0.20, significantly lower than the value 
obtained by most theoretical estimates. However, 
$\beta_2$ is not an observable and  its value 
depends on the model used to relate it to the transition moments. 
The formula used in Ref.~\cite{Dew03} contains a term coming 
from hexadecapole deformations which is usually not introduced.
The determination of $\beta_2$ from the experimental data and 
Eqn.~(\ref{beta2}) leads to a value compatible with the commonly 
estimated $\beta_2=0.27$.
%
%
\subsection{Monopole transition moments}
The $E0$ transition strength can be deduced from conversion-electron
measurements. The formulae which relate decay rates and
monopole strengths are a direct transposition to our model of 
the formulae given in Ref.\ \cite{Dan98}.

The nuclear electric monopole decay rate $T(E0)$ due to conversion 
electrons is given by~\cite{weneser}:
\begin{eqnarray}
\label{eq:te0ic}
T(E0) 
& = & 2.786 \cdot 10^{20} \, \rho^2_{E0} \frac{\Delta E}{2J+1} \,
      \bigg[ A(E0)_K + A(E0)_{L_I} 
      \nn \\
&   & + A(E0)_{L_{II}} + \ldots \bigg]
\end{eqnarray}
where $T(E0)$ is in $s^{-1}$ and the transition energy $\Delta E$ in MeV.
The electronic coefficients $A(E0)_i$, where $i$ represents the decay
channel, i.e.\ one of the electronic shells $K$, $L_I$, $L_{II}$, \ldots,
have been tabulated by Hager and Selzer~\cite{hager}.

The nuclear matrix element entering this
decay rate is the strength:
\begin{equation}
\rho_{E0}^2(J_{k'} \to J_k)
= \left| \frac{M(E0; J_{k'} \to J_k)}{R^2} \right|^2
\end{equation}
where $M(E0; J_{k'} \to J_k) = \langle J M k | r^2_p | J M k' \rangle$
and $R$ is $1.2 \, A^{1/3}$ fm.
%
%
\begin{figure}[t!]
\centerline{\epsfig{file=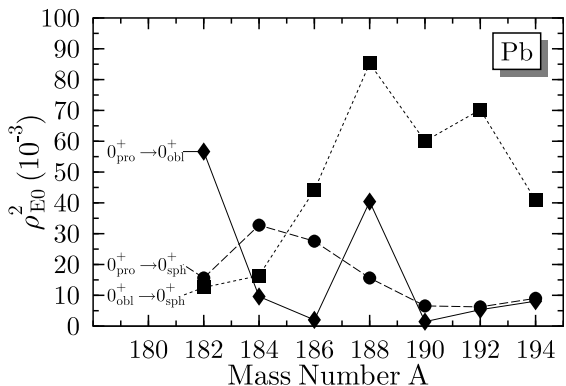}}
\caption{\label{fig:e0}
Monopole strength $\rho_{E0}^2$ for the transitions between
low-lying $0^+$ states. Available experimental data for 
the $0^+_2 \to 0^+_1$ transition in \nuc{190-194}{Pb}
are largely overestimated (see text). 
}
\end{figure}
%
%

The values calculated for the transitions between prolate, oblate
and spherical $0^+$ states are shown in Fig.\ \ref{fig:e0}. Their
systematics reflects the change of the potential landscapes that can 
be seen in Fig.\ \ref{fig:all:ejnz}. In most cases, the larger the
energy difference between the spherical and the deformed minima, 
the smaller the value of $\rho^2_{E0}$. The $\rho^2_{E0}$ values 
are quite small for transitions between prolate and oblate states, 
except for \nuc{188}{Pb} where the two states are very close in 
energy and  the oblate and prolate configurations are strongly mixed.

Experimental values for the monopole strength for the $(0^+_2 \to 0^+)$
transitions in \nuc{190-194}{Pb} are given in Ref.~\cite{Den89}:
$\rho_{E0}^2 \geq 6 \times 10^{-3}$ for \nuc{190}{Pb},
$\rho_{E0}^2 = (1.7 \pm 0.2) \times 10^{-3}$ for \nuc{192}{Pb}, and
$\rho_{E0}^2 = (1.0 \pm 0.2) \times 10^{-3}$ for \nuc{194}{Pb}.
Assuming that the observed $0^+_2$ states can be identified with our
oblate $0^+$ levels, our calculations overestimate 
these values by more than one order of magnitude for 
\nuc{192}{Pb} and \nuc{194}{Pb}. Interestingly,
transitions from the prolate $0^+$ states
to the ground state have the correct order magnitude.

If one estimates the mixing of configurations by the phenomenological
formula used in Ref.\ \cite{Den89}:
\begin{equation}
\label{rho_phen}
\rho^2_{E0}
= a^2 \, b^2 \left( \frac{3Z}{4\pi} \right)^2 \, \beta^4
,
\end{equation}
where $a$ and $b$ (with \mbox{$a^2 = 1-b^2$)} are the amplitudes of the 
spherical and oblate states in the physical wave functions, one obtains a 
mixing of around $10\%$ in our calculation, 20 to 30 times larger than 
the value deduced from the experimental data. Even though
Eqn.\ (\ref{rho_phen}) gives an extremely simple
and approximate picture of the configuration, it confirms that
our calculation probably predicts a much too large  mixing between 
the spherical and oblate configurations.

Ratios of $E0$ and $E2$ transition probabilities for \nuc{188}{Pb} have 
been experimentally determined by Dracoulis \etal\ \cite{Dra03}. 
These data are for transitions between states of the oblate and 
prolate bands  with the same angular momentum.
For \mbox{$J=2$}, they obtain a ratio equal to 2.6, to be compared
with 9.2 in our calculation, 2.2 for \mbox{$J=4$}, 1.2 for
\mbox{$J=6$} and 0.3 for \mbox{$J=8$}, to be compared with our calculated 
values of 0.27, 0.005 and 0.002, respectively. 

Using a simple band-mixing model adjusted 
using the experimental data, Dracoulis \etal\ \cite{Dra03} have 
determined the widths of all the $E0$ and $E2$ transitions depopulating
the oblate band. Our model also allows to determine these widths, 
without any additional assumptions. The widths of the dominant 
$E2$ transitions differ by at most an order of magnitude;
differences are  larger for $E0$ transitions, where our calculated 
values are lower by a factor varying from 10 for \mbox{$J=2$} to
500 for \mbox{$J=6$}. In the model of Dracoulis \etal, 
all states result from the mixing of a prolate and an oblate 
configuration only, while the mixing is much more complicated
in our calculation. The lower value for the $E0$ width that 
we obtain is probably related to larger differences in structure 
between the oblate and prolate bands in
the present calculation than in schematic models. 
The quadrupole transitions are in better agreement with the data.
The widths that can be calculated from the data of Dewald 
\etal\ \cite{Dew03}  agree with our computed values within a factor 2.
%
%
\section{Superdeformed structures}
Superdeformed bands have been extensively studied utilizing the 
cranked HFB method (see Ref.~\cite{BHR03} and references therein).
In this approach, the variation of the moment of inertia of a rotational band 
is studied up to high spins by means of mean-field states optimized for
each angular momentum with an appropriate cranking constraint. 
The variational space of the CHFB method is larger than in the
present calculations due to the breaking of time-reversal invariance.
As long as rotation takes place in a potential well which does not vary
too much with rotation, the cranking approximation is rather well justified.
Because of this feature, it is also well suited to describe the
moment of inertia of SD bands.

To challenge CHFB calculations, our method requires to project cranked 
HFB states on angular momentum; such a generalization is underway.
Nevertheless, we have extended our calculations to deformations 
at which superdeformed rotational bands have been observed in 
\nuc{192-196}{Pb}. Figure \ref{fig:all:ex1} shows the spectra obtained 
in the SD well after symmetry restorations and configuration mixing.
Compared to experiment, and to our previous CHFB calculation~\cite{Ter95}, 
the SD spectra are too compressed and the moments of inertia too large.
As the interaction that we used in Ref.~\cite{Ter95}  was the SLy4 Skyrme 
parameterization, we checked that the SLy6 parameterization used here 
leads to similar rotational bands 
when used in CHFB calculations and that agreement
with the experimental data is equally good.
%
%
\begin{figure}[t!]
\centerline{\epsfig{file=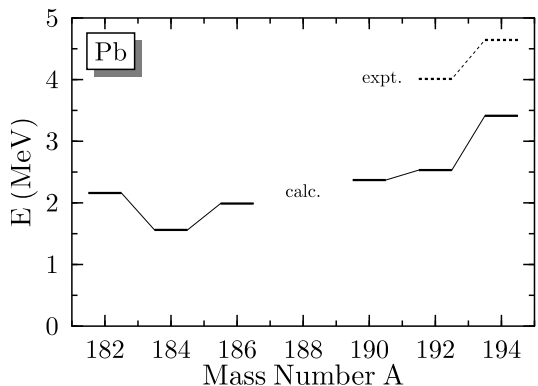}}
\caption{\label{fig:all:sd:bands}
Excitation energy of the $0^+$ bandheands of the
superdeformed rotational bands in \nuc{182-194}{Pb}.
}
\end{figure}
%
%
Without cranking,  our results for SD 
spectra cannot provide an accurate quantitative comparison with experiment.
However, our present method is very well adapted to determine the excitation 
energy of the SD well.
As a matter of fact, it can be viewed as a generalization of the method 
that we have already applied to the same problem~\cite{Mey95}
as correlations due to the restoration of angular momentum are included.
Looking at Fig.~\ref{fig:all:ex1}, one sees that bands are very 
well localized in the superdeformed well for \nuc{190-194}{Pb}.
Quadrupole deformations are slightly larger than in our cranking studies, 
with typical values of $\beta_2$ around 0.70, compared to 0.67 in CHFB 
for SLy4 and SLy6.

Transitions linking the SD bands to the normal-deformed well are known 
in \nuc{194}{Pb}~\cite{Lop96,Kru97} and, recently, in 
\nuc{192}{Pb}~\cite{Wil03}.
From these transitions one can estimate the ``experimental'' position 
of the $0^+$ bandheads in these two nuclei.
Calculated excitation energies of 2.5~MeV for \nuc{192}{Pb}
and 3.4~MeV for \nuc{194}{Pb}, are below the extrapolated 
experimental values of 4.425~MeV for \nuc{192}{Pb} and 4.878~MeV 
for \nuc{194}{Pb}.
We have repeated the same calculation with the SLy4 interaction 
used in a systematic study of the excitation energies of SD bands
\cite{Hee98} and obtained 4.7~MeV for \nuc{194}{Pb}.
This value is in much better agreement with the data.
This difference between both interactions gives a measure 
of the uncertainty related to the choice of the effective interaction. 
%
%
\begin{figure}[t!]
\centerline{\epsfig{file=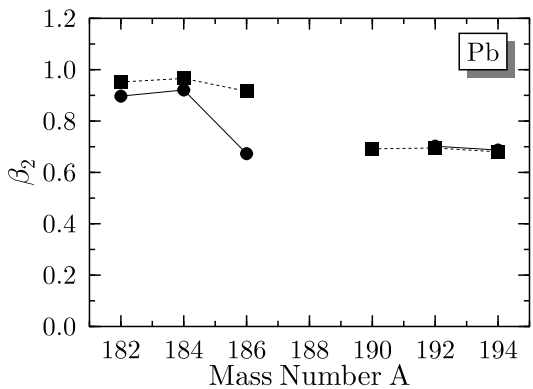}}
\caption{\label{fig:all:sd:beta}
Deformation of the SD states. squares: $\beta_2^{(s)}$ of the $2^+$ 
states, circles: $\beta_2^{(t)}(6^+ \to 4^+)$ (for \nuc{190-194}{Pb},
they lie on top of each other).
}
\end{figure}
%
%

A detailed analysis shows that the excitation energy of the SD band 
is substantially modified by the correlations. 
For \nuc{194}{Pb} for instance, the excitation energy of the
superdeformed minimum in the potential energy surface is lowered 
by angular momentum projection from 3.9~MeV to 3.0~MeV,
see \ Figs.~\ref{fig:all:enz} and \ref{fig:all:ejnz}. But 
after configuration mixing, the superdeformed $0^+$ state is pushed 
up again to 3.41~MeV. As can be seen in Fig.~\ref{fig:all:ex1},
quadrupole vibrations lower the SD state even further than the
projection, but this effect is larger for 
the ground state, so the SD bandheads are actually pushed 
toward larger excitation energy. For \nuc{192}{Pb}, angular momentum 
projection reduces the excitation energy from 3.0~MeV to 2.0~MeV, 
while the quadrupole correlations increase it back to 2.53~MeV.
In the same way, using the SLy4 interaction, the excitation energy 
of the SD band in \nuc{194}{Pb} is first reduced from 5.2~MeV to 
4.3~MeV and then increased back to 4.71~MeV by the GCM. A similar 
reduction of the excitation energy of SD bandheads in spherical 
nuclei was obtained for \nuc{16}{O} \cite{Ben03} and \nuc{40}{Ca} 
\cite{BFH03}. In those light nuclei, the absolute change of the 
excitation energy is more pronounced. 

No stable states are obtained in the SD well of \nuc{188}{Pb}.
For lighter isotopes, well-defined states are obtained again at
large deformation. They form nice rotational bands 
for \nuc{184}{Pb} and \nuc{182}{Pb}. Their excitation energies 
are quite small, below  3~MeV, see Fig.~\ref{fig:all:ex1}, 
even if one takes the uncertainty due to the interaction into account.
Their quadrupole moment is also significantly larger than for the known 
SD bands, leading to very strong $B(E2)$ transition probabilities 
within the bands.
%
%
\section{Discussion and Summary}
The main aim of this paper was to determine whether it is possible
to describe the coexistence of several low lying $0^+$ states
in neutron-deficient Pb isotopes by introducing correlations.
 In this respect, our results are very positive.
The evolution of the Pb  spectra with neutron number is 
qualitatively well reproduced. The direct link between  symmetry restored
wave functions and intrinsic mean-field configurations allows to
assign a predominantly oblate structure to the first excited $0^+$ state 
in the heaviest Pb isotopes and a predominantly prolate one
in the lightest ones. This result was not evident from a-priori analyses of
the mean-field deformation energy curves. In several cases, the
deformed minima are located in a well not expected to be sufficiently
deep to guarantee the
localization of a quantum state. Our results, obtained from
a purely quantum mechanical method based on an effective interaction as
the sole phenomenological ingredient, do support the intuitive
shape coexistence picture.

More detailed comparisons with the data show some discrepancies.
While the energies of the predominantly prolate states reproduce
the experimental data qualitatively, the energies of the oblate ones
are systematically overestimated and the spreading of the rotational bands
is too large. There could be different origins for these discrepancies.
First, the lack of triaxial deformations could induce an
overestimation of the energy of the oblate configuration. Their 
introduction should increase
the coupling between prolate and oblate states which seems to be too weak
when looking into $E0$ transition probabilities in \nuc{188}{Pb}. 
Second, mean-field states are generated by static calculations. 
This is not satisfactory from a variational point of view. As is well 
known, the projection of cranked states obtained for \mbox{$J_x=J$} is better
variationally and is expected to decrease the spreading of the bands. Such
an extension of our method is by far not trivial, but is presently underway. 

Finally, the effective interactions have not been constructed
to go beyond a mean-field approach. In principle, they should be readjusted, 
at least to account for the extra binding energies due to rotational and
vibrational correlations. Furthermore, our results could 
be affected by considering 
other generalizations of the density dependence of the Skyrme force
when calculating non-diagonal matrix elements, as suggested by Duguet 
and Bonche \cite{Dug03b,Dug03c}. Such an investigation is presently underway.
Even at the mean-field level, deficiencies of the Skyrme interactions
have been put into evidence. The analysis of the 
single-particle spectra in \nuc{208}{Pb} and \nuc{249}{Bk} matches up with the
conclusion drawn here from the neutron-number dependence of the oblate
configuration that the proton $1i_{13/2^+}$ shell is calculated
too high in excitation energy, above
the $1h_{9/2^-}$ level. This is a feature common to many effective
interactions which should be taken into account in a new fit of a force.

One of the major interests of our method is the determination of 
transition probabilities directly in the laboratory system, without
relying on ad hoc approximations. Unfortunately, the experimental data 
are very limited. The example of \nuc{188}{Pb} has shown how instructive the
comparison between theory and experiment for $E0$ and $E2$ transition 
probabilities can be. Experimental data for other isotopes are highly 
desirable.

Finally, we have extended our calculations up to quadrupole
deformations associated with superdeformed bands. As expected, 
for the isotopes in which these bands are known, the excitation 
energies are in reasonable agreement with experiment, but the 
moments of inertia are too large and not as good as in our previous 
cranking calculations. While the disappearance of SD structures is 
confirmed in \nuc{188-190}{Pb}, still more deformed configurations are 
predicted in the lightest isotopes. The experimental confirmation 
of this prediction is certainly a difficult experimental challenge, 
but would be an exciting result.
%
%
\section*{Acknowledgements}   
This research was supported in part by the \mbox{PAI-P5-07} of the
Belgian Office for Scientific Policy, and by the US Department of Energy
(DOE), Nuclear Physics Division, under contract no.\ W-31-109-ENG-38.
M.~B.\ acknowledges support through a European Community Marie Curie 
Fellowship. We thank G. Dracoulis, K.~Heyde, M.~Huyse, R.~V.~F.\ Janssens, R.\ Julin, 
G.\ Neyens, and P.\ Van Duppen for fruitful discussions. We acknowledge
the hospitality of ECT$^\ast$, the European Centre for Theoretical 
Studies in Nuclear Physics and Related Areas, Trento, where this work 
was finalized.
%
%


\begin{thebibliography}{99}

\bibitem{Jul01}
  R. Julin, K. Helariutta and M. Muikki,
  J. Phys. G \textbf{27}, R109 (2001).
 
\bibitem{Dup84}
  P. Van Duppen, E. Coenen, K. Deneffe, M. Huyse, K. Heyde, and 
  P. Van Isacker,
  Phys. Rev. Lett. \textbf{52}, 1974 (1984).

\bibitem{Dup87}
  P. Van Duppen, E. Coenen, K. Deneffe, M. Huyse, and J. L. Wood,
  Phys. Rev. C \textbf{35}, 1861 (1987).

\bibitem{And00}
  A. N. Andreyev, M. Huyse, P. Van Duppen, L. Weissman, D. Ackermann,
  J. Gerl, F. He{\ss}berger, S. Hofmann, A. Kleinbohl, G. M{\"u}nzenberg,
  S. Reshitko, C. Schlegel, H. Schaffner, P. Cagarda, M. Matos, S. Saro,
  A. Keenan, C. J. Moore, C. D. O'Leary, R. D. Page, M. J. Taylor, H. Kettunen,
  M. Leino, A. Lavrentiev, R. Wyss, and K. Heyde,
  Nature \textbf{405}, 430 (2000);
  Nucl. Phys. \textbf{A682}, 482c (2001).

\bibitem{Wau94}
  J. Wauters, N. Bijnens, H. Folger, M. Huyse, Han Yull Hwang,
  R. Kirchner, J. von Schwarzenberg, and P. Van Duppen,
  Phys. Rev. C \textbf{50}, 2768 (1994).

\bibitem{Woo92}
  J. L. Wood, K. Heyde, W. Nazarewicz, M. Huyse and P. Van Duppen,
  Phys. Rep. \textbf{215}, 101 (1992).


\bibitem{Hey83}
  K. Heyde, P. Van Isacker, M. Waroquier, J. L. Wood, and R. A. Meyer, 
  Phys. Rep. \textbf{102}, 291 (1983);

\bibitem{May77} 
  F. R. May, V. V. Pashkevich, and S. Frauendorf, 
  Phys. Lett. \textbf{68B}, 113 (1977).

\bibitem{Ben89}
  R. Bengtsson and W. Nazarewicz, 
  Z. Phys. \textbf{A334}, 269 (1989).

\bibitem{Sat91}
  W. Satu{\l}a, S. {\'C}wiok, W. Nazarewicz, R. Wyss, and A. Johnson, 
  Nucl. Phys. \textbf{A529}, 289 (1991).

\bibitem{Naz93}
  W. Nazarewicz,
  Phys. Lett. \textbf{B305}, 195 (1993).

\bibitem{Taj93}
  N. Tajima, H. Flocard, P. Bonche, J. Dobaczewski and P.-H. Heenen,
  Nucl. Phys. \textbf{A551} (1993) 409.

\bibitem{Lib99}
  J. Libert, M. Girod, and J.-P. Delaroche,
  Phys. Rev. C \textbf{60}, 054301 (1999).

\bibitem{Cha01}
  R. R. Chasman, J. L. Egido and L. M. Robledo,
  Phys. Lett. \textbf{B513}, 325 (2001).

\bibitem{Nik02}
  T. Niksic, D. Vretenar, P. Ring, and G. A. Lalazissis,
  Phys. Rev. C \textbf{65}, 054320 (2002).
 
\bibitem{Hey91}
   K. Heyde, J. Schietse, and C. De Coster,
  Phys. Rev. C \textbf{44}, 2216 (1991).

\bibitem{Cos00}
   C. De Coster, B. Decroix, and K. Heyde,
   Phys. Rev. C \textbf{61}, 067306 (2000).

\bibitem{Fos03}
  R. Fossion, K. Heyde, G. Thiamova, and P. Van Isacker,
  Phys. Rev. C \textbf{67}, 024306 (2003). 

\bibitem{Dup90}
  P. Van Duppen, M. Huyse and J. L. Wood,
  J. Phys. \textbf{G16}, 441 (1990).

\bibitem{Del96}
  D. S. Delion, A. Florescu, M. Huyse, J. Wauters, P. Van Duppen, A. Insolia, 
  and R. J. Liotta,
  Phys. Rev. C \textbf{54}, 1169 (1996)  

\bibitem{All98}
  R. G. Allatt, R. D. Page, M. Leino, T. Enqvist, K. Eskola, P. T. Greenlees, 
  P. Jones, R. Julin, P. Kuusiniemi, W. H. Trzaska and J. Uusitalo,
  Phys. Lett. \textbf{B437}, 29 (1998).

\bibitem{Hey88}
  K. Heyde and R. A. Meyer, 
  Phys. Rev. C \textbf{37}, 2170 (1988).

\bibitem{Val00}
  A. Valor, P.-H. Heenen and P. Bonche,
  Nucl. Phys. \textbf{A671}, 145 (2000).

\bibitem{Ben03}
  M. Bender and P.-H. Heenen,
  Nucl. Phys. \textbf{A713}, 390 (2003).

\bibitem{Dug03a}
  T. Duguet, M. Bender, P. Bonche, P.-H. Heenen,
  Phys. Lett. \textbf{B559}, 201 (2003).

\bibitem{Fle03}
  P. Fleischer, P. Kl{\"u}pfel, M. Bender, and P.-G. Reinhard,
  preprint 2003.

\bibitem{Hil53}
  D. L. Hill and J. A. Wheeler,
  Phys. Rev. \textbf{89}, 1102 (1953);
  J. J. Griffin and J. A. Wheeler, 
  Phys. Rev. \textbf{108}, 311 (1957).

\bibitem{Cha98}
  E. Chabanat, P. Bonche, P. Haensel, J. Meyer, and R. Schaeffer,
  Nucl. Phys. \textbf{A635}, 231 (1998);
  Nucl. Phys. \textbf{A643},441(E) (1998).

\bibitem{Rig99}
  C. Rigollet, P. Bonche, H. Flocard, P.-H. Heenen,
  Phys. Rev. C \textbf{59}, 3120 (1999).
 
\bibitem{BFH03}
  M. Bender, H. Flocard and P.-H. Heenen,
  Phys. Rev. C \textbf{68}, 044321 (2003).

\bibitem{Cwi00}  
  S. {\'C}wiok,
  private communication.


\bibitem{Lan95}
  G. J. Lane, G. D. Dracoulis, A. P. Byrne, P. M. Walker, A. M. Baxter, 
  J. A. Sheikh and W. Nazarewicz,
  Nucl. Phys. \textbf{A586}, 316 (1995).

\bibitem{Vyv02}
  K. Vyvey, A. M. Oros-Peusquens, G. Neyens, D. L. Balabanski, D. Borremans, 
  S. Chmel, N. Coulier, R. Coussement, G. Georgiev, H. H{\"u}bel,
  N. Nenoff, D. Rossbach, S. Teughels and K. Heyde,
  Phys. Lett. \textbf{B538}, 33 (2002);
  K. Vyvey, D. Borremans, N. Coulier, R. Coussement, G. Georgiev, S. Teughels,
  G. Neyens, H. H{\"u}bel, and D. L. Balabanski,
  Phys. Rev. C \textbf{65}, 024320 (2002);
  K. Vyvey, S. Chmel, G. Neyens, H. H{\"u}bel, D. L. Balabanski, 
  D. Borremans, N. Coulier, R. Coussement, G. Georgiev, N. Nenoff, 
  S. Pancholi, D. Rossbach, R. Schwengner, S. Teughels, and S. Frauendorf,
  Phys. Rev. Lett. \textbf{88}, 102502 (2002).

\bibitem{Smi03a}
  N. A. Smirnova, P.-H. Heenen, and G. Neyens,
  Phys. Lett. \textbf{B569}, 151 (2003).

\bibitem{BBDH03}
  M. Bender, P. Bonche, T. Duguet, and P.-H. Heenen,
  Nucl. Phys. \textbf{A723}, 354 (2003). 

\bibitem{Ben99} 
  M. Bender, K. Rutz, P.-G. Reinhard, J. A. Maruhn, and W. Greiner, 
  Phys. Rev. C \textbf{60}, 034304 (1999).

\bibitem{Kle02}
  M. Kleban, B. Nerlo-Pomorska, J. F. Berger, J. Decharg{\'e}, 
  M. Girod, and S. Hilaire,
  Phys. Rev. C \textbf{65}, 024309 (2002).

\bibitem{Dra03}
  G. D. Dracoulis, G. J. Lane, A. P. Byrne, A. M. Baxter, T. Kib{\'e}di, 
  A. O. Macchiavelli, P. Fallon, and R. M. Clark,
  Phys. Rev. C \textbf{67}, 051301(R) (2003).

\bibitem{Dew03}
  A. Dewald, R. Peusquens, B. Saha, P. von Brentano, A. Fitzler, T. Klug, 
  I. Wiedenh{\"o}ver, M. P. Carpenter, A. Heinz, R. V. F. Janssens, 
  F. G. Kondev, C. J. Lister, D. Seweryniak, K. Abu Saleem, R. Kr{\"u}cken, 
  J. R. Cooper, C. J. Barton, K. Zyromski, C. W. Beausang, Z. Wang, 
  P. Petkov, A. M. Oros-Peusquens, U. Garg, and S. Zhu,
  Phys. Rev. C \textbf{68}, 034314 (2003).

\bibitem{Hee01}
  P.-H. Heenen, A. Valor, M. Bender, P. Bonche, and H. Flocard,
  Eur. Phys. J. \textbf{A11}, 393 (2001).

\bibitem{Dan98}
  H. Dancer, P. Bonche, H. Flocard, P.-H. Heenen, J. Meyer, and M. Meyer,
  Phys. Rev. C \textbf{58}, 2068 (1998).

\bibitem {weneser}
  E. L. Church and J. Weneser, 
  Phys. Rev. \textbf{103}, 1035 (1956).

\bibitem{hager}
  R. S. Hager and E. C. Seltzer, 
  Nucl. Data Tables \textbf{A6}, 1 (1969).

\bibitem{Den89}
  P. Dendooven, P. Decrock, M. Huyse, G. Reusen, P. Van Duppen and J. Wauters,
  Phys. Lett. \textbf{B226}, 27 (1989).

\bibitem{BHR03}
  M. Bender, P.-H. Heenen and P.-G. Reinhard,
  Rev. Mod. Phys. \textbf{75}, 121 (2003).

\bibitem{Ter95}
  J. Terasaki, P.-H. Heenen, P. Bonche, J. Dobaczewski, and H. Flocard,
  Nucl. Phys. \textbf{A593}, 1 (1995); \\
  P. Fallon, P.-H. Heenen, W.Satu{\l}a, R. M. Clark, F. S. Stephens, 
  M. A. Deleplanque, R. M. Diamond, I. Y. Lee, A. O. Macchiavelli, 
  and K.Vetter,
  Phys. Rev. C \textbf{60}, 044301 (1999).

\bibitem{Mey95}
  J. Meyer, P. Bonche, M. S. Weiss, J. Dobaczewski, H. Flocard 
  and P.-H. Heenen,
  Nucl. Phys. \textbf{A588}, 597 (1995).

\bibitem{Lop96}
  A. Lopez-Martens, F. Hannachi, A. Korichi, C. Sch{\"u}ck, E. Gueorguieva, 
  Ch. Vieu, B. Haas, R. Lucas, A. Astier, G. Baldsiefen, M. Carpenter, 
  G. de France, R. Duffait, L. Ducroux, Y. Le Coz, Ch. Finck, A. Gorgen, 
  H. H{\"u}bel, T. L. Khoo, T. Lauritsen, M. Meyer, D. Pr{\'e}vost, 
  N. Redon, C. Rigollet, H. Savajols, J. F. Sharpey-Schafer, 
  O. Stezowski, Ch. Theisen, U. Van Severen, J. P. Vivien 
  and A. N. Wilson,
  Phys. Lett. \textbf{B380}, 18 (1996).

\bibitem{Kru97}
  R. Kr{\"u}cken, S. J. Asztalos, J. A. Becker, B. Busse, R. M. Clark, 
  M. A. Deleplanque, A. Dewald, R. M. Diamond, P. Fallon, K. Hauschild, 
  I. Y. Lee, A. O. Macchiavelli, R. W. MacLeod, R. Peusquens, G. J. Schmid, 
  F. S. Stephens, K. Vetter, and P. von Brentano,
  Phys. Rev. C \textbf{55}, R1625 (1997).

\bibitem{Wil03}
  A. N. Wilson, G. D. Dracoulis, A. P. Byrne, P. M. Davidson, 
  G. J. Lane, R. M. Clark, P. Fallon, A. G{\"o}rgen, 
  A. O. Macchiavelli, and D. Ward,
  Phys. Rev. Lett. \textbf{90}, 142501 (2003).

\bibitem{Hee98}
  P.-H. Heenen, J. Dobaczewski, W. Nazarewicz, P. Bonche and T. L. Khoo,
  Phys. Rev. C \textbf{57}, 1719 (1998).

\bibitem{Bon90}
  P. Bonche, J. Dobaczewski, H. Flocard, P.-H. Heenen and J. Meyer,
  Nucl. Phys. \textbf{A510}, 466 (1990).

\bibitem{Dug03b}
  T. Duguet, 
  Phys. Rev. C \textbf{67}, 044311 (2003).

\bibitem{Dug03c}
  T. Duguet and P. Bonche,
  Phys. Rev. C \textbf{67}, 054308 (2003).

\end{thebibliography}
\end{document}